\documentclass[preprint,aps,amsmath,nofootinbib,tightenlines]{revtex4}
\usepackage{graphicx}
\usepackage{bm}
\usepackage{epsfig}

\def\Dsl{\hbox{/\kern-.6000em D}} 

\def\dsl{\,\raise.15ex\hbox{/}\mkern-13.5mu D}

\def\ltap{\ \raise.3ex\hbox{$<$\kern-.75em\lower1ex\hbox{$\sim$}}\ }
\def\gtap{\ \raise.3ex\hbox{$>$\kern-.75em\lower1ex\hbox{$\sim$}}\ }
\def\OMIT#1{}

\def\lsim{\mathrel{\raise.3ex\hbox{$<$\kern-.75em\lower1ex\hbox{$\sim$}}}}
\def\gsim{\mathrel{\raise.3ex\hbox{$>$\kern-.75em\lower1ex\hbox{$\sim$}}}}

\newcommand{\bmp}{\mathbf p}

\catcode`\@=11
\def\slash{\mathpalette\make@slash}
\def\make@slash#1#2{\setbox\z@\hbox{$#1#2$}%
  \hbox to 0pt{\hss$#1/$\hss\kern-\wd0}\box0}
\catcode`\@=12 

\begin{document}


\preprint{ \vbox{ \hbox{MPP-2006-36} 
}}

\title{\phantom{x}\vspace{0.5cm} 
Next-to-leading-logarithmic QCD Corrections to \\[3mm]
the Cross Section $\sigma(e^+e^-\to t\bar t H)$ at $500$~GeV
\vspace{1.0cm} }

\author{Cailin Farrell and Andr\'e~H.~Hoang\vspace{0.5cm}}
\affiliation{Max-Planck-Institut f\"ur Physik\\
(Werner-Heisenberg-Institut), \\
F\"ohringer Ring 6,\\
80805 M\"unchen, Germany\vspace{1cm}
\footnote{Electronic address: ahoang@mppmu.mpg.de, farrell@mppmu.mpg.de}\vspace{1cm}}


\begin{abstract}
\vspace{0.5cm}
\setlength\baselineskip{18pt}
We determine the next-to-leading logarithmic (NLL) QCD corrections to the
cross section $\sigma(e^+e^-\to t\bar t H)$ for center-of-mass (c.m.)
energies up to $500$~GeV. The dynamics is dominated by nonrelativistic
effects, and the summation of terms singular in the relative $t\bar t$
velocity is mandatory to all orders in the strong coupling constant
$\alpha_s$ using an effective theory. The summations lead to an
enhancement of the tree level predictions by about a factor of two 
and are important for the determination of the top Yukawa coupling. We
also study the impact of polarization of the electron-positron beams
and provide a fast approximation formula for the known $\mathcal
O(\alpha_s)$  QCD fixed-order prediction. 
\end{abstract}
\maketitle


\newpage

%
%
%
\section{Introduction}
\label{sectionintroduction}

The discovery and exploration of the mechanism of mass generation and
electroweak symmetry breaking is one of the most important tasks of future
collider experiments. Within the Standard Model of elementary particle
physics (SM) electroweak symmetry breaking is realized by the Higgs
mechanism which postulates the existence of an electric neutral
elementary scalar field that interacts with all SM particles carrying
nonzero hypercharge and weak isospin. Through self-interactions this
Higgs field acquires a vacuum expectation value
$V=(\sqrt{2}\,G_F)^{1/2}\approx 246$~GeV, $G_F$ being the Fermi
constant, which breaks the SU(2)$_L\times$U(1)$_Y$ symmetry at high
energies down to the electric U(1)$_{\rm em}$ below the symmetry
breaking scale and leads to nonzero masses of the elementary
particles. The Higgs mechanism also predicts that the Higgs fields can
be produced as a massive Bose particle in collider experiments
if sufficient energy is provided in the process. The mass of the
Higgs boson is expected to lie between the current experimental lower
limit of $114.4$~GeV~\cite{LEPlimits} and about 1~TeV. Current
analyses of electroweak precision observables yield a $95\%$~CL upper
indirect bound of $186$~GeV for the Higgs boson
mass~\cite{MHupperlimit}. While a Higgs boson with a mass up to
$1$~TeV can be found at the LHC, precise and model-independent
measurements of quantum numbers and couplings are likely to be
restricted to a future $e^+e^-$ Linear
Collider~\cite{TESLATDR,ALCphysics,ACFALCphysics} such as the 
International Linear Collider (ILC) project. 

The Higgs mechanism predicts that the quark masses $m_q$ are related
to the quark-Higgs Yukawa coupling $\lambda_q$ through the relation
$m_q=\lambda_qV$. This makes the measurement of the Yukawa coupling to
the top quark ($m_t=172.5\pm 2.3$~GeV~\cite{topmass}) particularly important
since it is expected to have a high precision. At
a future $e^+e^-$ Linear Collider the top Yukawa coupling can be
measured from the process $e^+e^-\to t\bar t H$ since 
the amplitudes describing Higgs radiation off the $t\bar t$
pair dominate the cross section.~\footnote{An indirect measurement through virtual Higgs effects
might be also possible at the $t\bar t$ threshold if the Higgs mass is
close to the present lower experimental limit~\cite{TESLATDR}.}

For the second phase of the ILC project with c.m.\,energies between
$500$~GeV and $1$~TeV and assuming a Higgs mass of around $120$~GeV
the total cross section $\sigma(e^+e^-\to t\bar t H)$ is at the level
of $1-2$~fb and measurements of $\lambda_t$ with experimental errors
of around five percent are
expected~\cite{JustetopYukawa,GaytopYukawa}. The precision 
motivates the computation of radiative corrections. In the
approximation that the top 
quark and the Higgs boson are stable particles\footnote{
For a light Higgs boson this is an excellent approximation. 
For $m_H=115(150)$~GeV one finds
$\Gamma_H=0.003(0.017)$~GeV~\cite{Hdecay}}
the tree level cross section was determined already some time ago in
Refs.~\cite{Borneetth}. The full set of one-loop QCD corrections were
obtained in Ref.~\cite{Dittmaier1}. Earlier studies using
approximations were given in Refs.~\cite{Dawson1,Dawson2}. One-loop
electroweak corrections were studied in Refs.~\cite{Belanger1,Denner1}
and also in Ref.~\cite{You1}. 

%
%
\begin{figure}[t] 
\begin{center}
 \leavevmode
 \epsfxsize=9cm
 \leavevmode
 \epsffile[140 300 430 495]{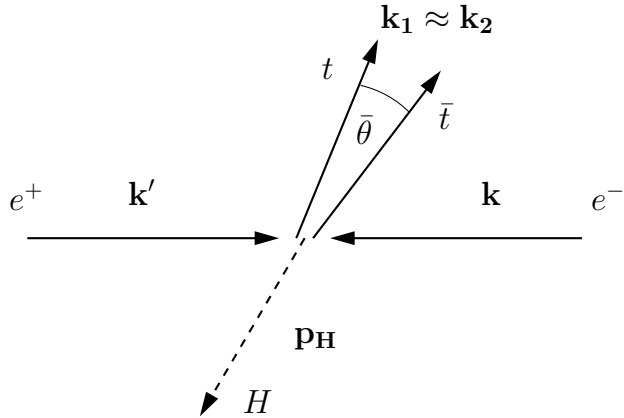}
 \vskip  0.0cm
 \caption{
Typical constellation of momenta for the process $e^+e^-\to t\bar t H$ in the
large Higgs energy endpoint region. 
 \label{fig1} }
\end{center}
\end{figure}

The phase space region where the Higgs energy is close to its upper
endpoint,
\begin{equation}
E_H\approx E_H^0 \equiv (s+m_H^2-4m_t^2)/(2\sqrt s)\;,
\end{equation}
$\sqrt s$ being the
center of mass energy, was studied in detail in
Ref.~\cite{FarrellHoang1}. In the
large Higgs energy endpoint region the $t\bar t$ pair is forced to
become collinear and to move opposite to the Higgs direction
in order to maximize the momentum necessary to balance the large Higgs
momentum, see Fig.~\ref{fig1}. Thus the $t\bar t$ invariant mass is
close to $2m_t$. In this kinematic regime the $t\bar t$ pair is
nonrelativistic in its c.m.\,frame and fixed-order QCD perturbation
theory in powers of $\alpha_s$ leads to singular terms proportional to
$(\alpha_s/v)^n$ and $(\alpha_s\ln v)^n$ which have to be summed to
all order. Here, $v=(1-4m_t^2/Q^2)^{1/2}$ is the top quark relative
velocity in the $t\bar t$ c.m.\,frame and $Q$ is the $t\bar t$
invariant mass. In Ref.~\cite{FarrellHoang1} these singularities were
summed at NLL order in a simultaneous expansion in $\alpha_s$ and $v$
and also accounting for the finite top quark width. The computations
were carried out using a nonrelativistic effective
theory~\cite{LMR,HoangStewartultra,hmst} originally developed for the 
threshold region in the process $e^+e^-\to t\bar t$. Due to the large
top quark width, $\Gamma_t\approx 1.5$~GeV, the nonrelativistic $t\bar
t$ dynamics is protected from nonperturbative effects and the
summations can be carried out with perturbative methods. It was shown
in Ref.~\cite{FarrellHoang1} that the summation of the singular terms
leads to an enhancement of the total cross section that needs to be
accounted for up to c.m.\,energies of about $700$~GeV. The impact of
the summations increases with the fraction of the phase space
where the c.m.\,top velocity $v$ is nonrelativistic, i.e. it increases with
the Higgs and top 
quark masses and decreases with the c.m.\ energy. A convenient measure
for the impact of the nonrelativistic summations on the total cross
section is the maximal relative velocity of the $t\bar t$ pair which is
achieved at the {\it low} Higgs energy endpoint $E_H=m_H$,
\begin{equation}
v^{\rm max} \, = \, \left(\,1-\frac{4m_t^2}{\;Q^2_{\rm max}}\right)^{1/2} \, 
= \, \left(\,1-\frac{4m_t^2}{(\sqrt{s}-m_H)^2}\,\right)^{1/2}
\,.
\label{vmaxdef}
\end{equation}
For small $v^{\rm max}$ the summations have a large effect since 
the available phase space is predominantly nonrelativistic.   

As was already demonstrated in Ref.~\cite{FarrellHoang1}, the
fixed-order QCD predictions~\cite{Dittmaier1,Dawson1,Dawson2} become
unreliable for c.m.\,energies up to 
$500$~GeV, which corresponds to the energy available during the first phase of
the ILC project. For $m_H=(120,130,140)$~GeV, $m_t=175$~GeV, and
$\sqrt{s}=500$~GeV one has $v^{\rm max}=(0.39,0.32,0.23)$ and consequently 
the entire phase space is governed by the
nonrelativistic QCD dynamics. The nonrelativistic expansion
based on the parametric counting $\alpha_s\sim v\ll 1$ has to be employed
rather then the $\alpha_s$ expansion to make reliable theoretical
predictions for the cross section. Another consequence of small
$v^{\rm max}$ is that the cross section for
c.m.~energies up to $500$~GeV can be substantially smaller than
$1$~fb due to phase space suppression, which severely restricts
statistics. Since the singularities proportional to $(\alpha_s/v)^n$
and $\alpha_s\ln v$ are large in this case only predictions where the
nonrelativistic summations are accounted for allow for a 
realistic assessment of Yukawa coupling measurements during the first phase of
the ILC project~\cite{Justetalk,topQCDSnowmass}.

In this work we give a detailed analysis of the total cross section
and the Higgs energy distribution for the process $e^+e^-\to t\bar t
H$ for c.m.\,energies up to $500$~GeV accounting for QCD effects at
NLL order in the nonrelativistic expansion. The approach
of Ref.~\cite{FarrellHoang1} developed for descriptions of the large
Higgs energy endpoint region is extended to the case
where the entire phase space is nonrelativistic. We show that our NLL order
predictions are substantially larger than the known tree level predictions,   
which have in fact been used for experimental simulations
studies at $500$~GeV in the past~\cite{Justetalk}. We also account for the
possibility of electron-positron 
beam polarization which can further enhance the cross section. Our results
significantly affect the prospects for top Yukawa coupling measurements 
during the first phase of the ILC project. 

The content of this paper is organized as follows:
In Sec.~\ref{sectionlargeHE} we review the ingredients of the
factorization formula derived in Ref.~\cite{FarrellHoang1} in the
large Higgs energy endpoint region valid for large c.m.\,energies. 
We extend the presentation by also accounting for electron-positron
beam polarization and by giving a more detailed discussion of the
$t\bar t$ final state in the helicity basis. In 
Sec.~\ref{sectionlowHE} we discuss the modifications that need to be
applied to the factorization formula for the case where the full
phase space is nonrelativistic. In Sec.~\ref{sectionanalysis} we analyze
our results numerically and 
Sec.~\ref{sectionconclusion} contains the conclusion.

\section{The Large Higgs Energy Endpoint Region} 
\label{sectionlargeHE}

In the large Higgs energy region $E_H\approx E_H^0$ the Higgs energy
distribution can be factorized into a hard part describing the production of
the $t\bar t$ pair and the Higgs boson and in a low-energy part describing the
nonrelativistic dynamical QCD effects of the $t\bar t$ subsystem. The
latter are responsible for the singularities proportional to powers of
$\alpha_s/v$ and $\alpha_s\ln v$. The factorization formula, valid at
NLL order for unpolarized electron-positron beams and top quarks, was
derived in Ref.~\cite{FarrellHoang1}. Accounting for electron-positron
beam polarization and polarized top quarks the factorization formula
for fully polarized electrons and positrons has the form
\begin{eqnarray}
\lefteqn{
\left(\frac{d\sigma}{d E_H}(E_H\approx E_H^{0})\right)^\pm \, = \,
\frac{8\,N_c\,\left[(1+x_H-4x_t)^2-4x_H\right]^{1/2}}{s^{3/2}\,m_t^2}\,
}
\nonumber\\[2mm]& &\mbox{} \hspace{1cm}
\times\,\left(\,c^2_0(\nu)\, F^Z_{0,\pm} + 
\sum_{i=-1,0,+1}c_{(1,i),\pm}^2(\nu)\,F_{(1,i),\pm}^{\gamma Z}\,\right)\,
\,\mbox{Im}\left[\, G^c(C_F\alpha_s(m_t \nu),v,m_t,\nu)\,\right]\,
\,,
\label{dsdEHEFT}
\end{eqnarray}
with
\begin{eqnarray}
x_t & \equiv & \frac{m_t^2}{s}\,,\qquad 
x_H \, \equiv \, \frac{m_H^2}{s}\,,\qquad
x_Z \, \equiv \, \frac{m_Z^2}{s}
\,.
\label{const2}
\end{eqnarray}
Here, $c_0$ and $c_{(1,i)}$ are the hard singlet and triplet
QCD Wilson coefficients which depend on the effective theory renormalization 
parameter $\nu$, $ F^Z_{0,\pm}$ and $F_{(1,i),\pm}^{\gamma Z}$ are the
hard electroweak tree-level matching conditions, and $G^c$ is the 
Green's function of the NLL Schr\"odinger equation of the effective
theory for the top quarks. A detailed discussion of these quantities
will follow shortly. 

The index denotes the helicity of the electrons, i.e. ``$-$'' refers
to right-handed positrons and left-handed electrons and the index
``$+$'' refers to left-handed positrons and right-handed
electrons. Since the electron mass is neglected, the cross section
vanishes if both electron and positron have the same helicity. 
For arbitrary polarization $P_+$ of the positrons and $P_-$ of the
electrons the spectrum reads 
\begin{eqnarray}
\left(\frac{d\sigma}{d E_H}\right)
& = & 
\frac{1}{4}(1+P_-)(1-P_+)\, \left(\frac{d\sigma}{d E_H}\right)^+
\,+\,\frac{1}{4}(1-P_-)(1+P_+)\, \left(\frac{d\sigma}{d E_H}\right)^-
\,,
\end{eqnarray}
where the polarization of a beam  with $N_+$ right-handed particles and
$N_-$ left-handed particles is defined as
\begin{equation}
P=\frac{N_+-N_-}{N_++N_-}\equiv\frac{N_+-N_-}{N_{\rm tot}}
\end{equation}
and can take on values between $-1$ and $1$.

The first two terms is Eq.~(\ref{dsdEHEFT}) are the hard factors and the third 
term is the imaginary part of the zero-distance Green function of the NLL
Schr\"odinger equation that can be derived from the effective theory
Lagrangian. The Green function describes the effects of the low-energy
nonrelativistic dynamics on the $t\bar t$ production rate for the top pair
being in an S-wave state and does not depend on the polarization of
the electron-positron beams. It depends on the effective theory
renormalization scaling parameter $\nu$ and is proportional to the
time-ordered product of the effective theory operators describing the
nonrelativistic QCD dynamics for the
production and annihilation of the $t\bar t$ pair at leading
logarithmic (LL) and NLL order.~\footnote{
The renormalization scaling parameter $\nu$ has mass dimension zero and is
used in the effective theory to describe the correlated running 
of soft and ultrasoft fluctuations~\cite{LMR}. The hard effective
theory matching scale (at the top quark mass) is at $\nu=1$ and 
low-energy matrix elements are evaluated for $\nu\sim v\sim \alpha_s$ to
avoid the appearance of large logarithmic terms.
} 
At LL
order (in dimensional regularization)
the Green function has the simple analytic form 
\begin{eqnarray}
 G^c_{\rm LL}(a,v,m_t,\nu) & = &
 \frac{m_t^2}{4\pi}\left\{\,
 i\,v - a\left[\,\ln\left(\frac{-i\,v}{\nu}\right)
 -\frac{1}{2}+\ln 2+\gamma_E+\psi\left(1\!-\!\frac{i\,a}{2\,v}\right)\,\right]
 \,\right\}
 + \,\frac{m_t^2\,a}{4 \pi}\,\,\frac{1}{4\,\epsilon}
\,.
\nonumber\\
\label{deltaGCoul}
\end{eqnarray}  
For the NLL order Green function we use the numerical techniques and codes of
the  TOPPIC program developed in Ref.~\cite{Jezabek1} (see also
Ref.~\cite{Strassler1}) and determine an exact solution of the full NLL
Schr\"odinger equation employing the approach of Refs.~\cite{hmst}. 
We estimate the QCD uncertainties in the normalization of the Higgs energy
spectrum from the NLL order Green function as 5\%~\cite{FarrellHoang1,HoangEpi}.
Note that
we account for the top quark finite lifetime by shifting the $t\bar t$
invariant mass $Q$ used in the Green function into the complex plane such
that the top quark relative velocity reads
\begin{eqnarray}
v & = &
\sqrt{\frac{Q-2 m_t-2\delta m_t(\nu)+i\Gamma_t}{m_t}}
\,,
\label{vdef}
\end{eqnarray}
where
\begin{eqnarray}
Q^2 \, = \, s + m_H^2 - 2\sqrt{s}\, E_H
\,.
\end{eqnarray}
This accounts for the top quark finite lifetime consistently at LL
order, see for example~\cite{HoangReisser1}. A consistent NLL description of 
finite lifetime effects and electroweak corrections shall be included in a
subsequent 
publication. The term $\delta m_t$ in Eq.~(\ref{vdef}) is a residual mass term
that has to be specified perturbatively at each order to fix which top quark
mass definition is being employed. In the pole mass scheme the residual
mass term vanishes to all orders. We use the 1S mass
scheme~\cite{Hoangupsilon,HoangTeubner}. The corresponding expression for
$\delta m_t$ at NLL order 
can also be found in Ref.~\cite{FarrellHoang1}. We use the 1S top quark mass and
implement the residual mass term in the soft factor of the factorization
formula because it avoids the pole mass renormalon problem~\cite{Vcrenormalon}
and leads to a $t\bar t$ resonance peak position that is stable under higher
order perturbative corrections~\cite{synopsis}. For the NLL order QCD
corrections to the hard factors, which are discussed below, we neglect the
corrections that arise from the residual mass term because the numerical
effects are at the 1\% level and substantially smaller than the uncertainties
from low-energy QCD effects. This approximation was also used in
Ref.~\cite{FarrellHoang1}.

Concerning the hard contributions in Eq.~(\ref{dsdEHEFT}),
the first term in the parenthesis gives the
contribution for the $t\bar t$ pair in a S-wave spin singlet state and the other
three terms give the contributions for the $t\bar t$ pair in the three 
S-wave spin triplet 
$(+1,0,-1)$ states. As described already in Ref.~\cite{FarrellHoang1} we use
the helicity basis for the top and antitop spinors in the endpoint where
$k_1=k_2$ (see Fig.~\ref{fig1}) to define the singlet and the triplet states.  
In this basis there are additional $v$-suppressed (NLL) contributions to the
triplet $\pm 1$ contribution that arise from S-P wave interference
terms and originate from the interference of vector and axial-vector
contributions at the $t\bar t$ vertex. These additional order $v$
contributions cancel in the sum of the 
triplet contributions and can also be avoided if a spin basis is used that
does not depend on the momenta of the top quarks~\cite{Parke1}. Since here we
are not interested in the phenomenology of top polarization these additional
NLL order contributions are not included in Eq.~(\ref{dsdEHEFT}).
The functions $F^{Z,\gamma Z}$ are the tree level (hard) matching conditions
for the contributions of the respective $t\bar t$ spin states. They read
\begin{eqnarray}
F_{(1,+1),\pm}^{\gamma,Z} & = & F_{(1,-1),\pm}^{\gamma,Z}
\nonumber\\[2mm] & = & 
\frac{ 2\alpha^2 \lambda_t^2}{6}\, 
\frac{(1 - x_H + 4x_t)^2}{(1 + x_H - 4x_t)^2}\,
\left (\, Q_e^2 Q_t^2  + \frac{v_t^2\left(v_e \mp a_e\right)^2}{(1 - x_Z)^2} 
    + \frac{2 Q_e Q_t \left(v_e \mp a_e\right) v_t}{(1 - x_Z)} 
\right)
\nonumber\\[2mm] & &
 + \, \frac{4 \alpha^2 g_Z \lambda_t}{3}\,
\frac{(x_t x_Z)^{1/2}(1 - x_H + 4x_t)}
{(1 + x_H - 4x_t)(4x_t - x_Z)(1 - x_Z)}\,
\left( \frac{v_t^2 \left(v_e \mp a_e\right)^2}{(1 - x_Z)} 
   + Q_e  Q_t (v_e \mp a_e)v_t \right)
\nonumber\\[2mm] & &
 +  \frac{4\alpha^2 g_Z^2  v_t^2\left(v_e \mp a_e\right)^2}{3}\,
  \frac{x_t x_Z}{(4x_t - x_Z)^2(1 - x_Z)^2}
\,,
\label{F1def}
\end{eqnarray}
\begin{eqnarray}
F_{(1,0),\pm}^{\gamma,Z} & = &
\frac{16 \alpha^2 \lambda_t^2}{3}\,
 \frac{ x_t}{(1 + x_H - 4x_t)^2}\,
 \left(\, 
    Q_e^2 Q_t^2  + \frac{v_t^2\left(v_e \mp a_e\right)^2}{(1 - x_Z)^2} 
     + \frac{2 Q_e Q_t \left(v_e \mp a_e\right) v_t}{(1 - x_Z)} 
\right) 
\nonumber\\[2mm] & &
 + \frac{4 \alpha^2 g_Z \lambda_t}{3} \,
\frac{(x_t x_Z)^{1/2}(1 - x_H + 4x_t)}
{(1 + x_H - 4x_t)(4x_t - x_Z)(1 - x_Z)}\,
\left( \frac{v_t^2 \left(v_e \mp a_e\right)^2}{(1 - x_Z)}
   + Q_e Q_t \left(v_e \mp a_e\right) v_t 
\right)
\nonumber\\[2mm] & & +
  \frac{ \alpha^2 g_Z^2 v_t^2\left(v_e \mp a_e\right)^2}{12}\,
  \frac{(1 - x_H + 4x_t)^2 x_Z}{(4x_t - x_Z)^2(1 - x_Z)^2} 
\,,
\\[4mm]
F_{0,\pm}^{Z} & = &
\frac{\alpha^2 g_Z^2 a_t^2\left(v_e \mp a_e\right)^2}{12}\,
\frac{(1 - x_H + 4x_t)^2 - 16 x_t}{(1 - x_Z)^2\,x_Z}
\,,
\label{F0def}
\end{eqnarray}
where
\begin{eqnarray}
  v_f = \frac{T_3^f-2 Q_f s_w^2}{2s_w c_w}\,,
  \quad
  a_f = \frac{T_3^f}{2s_w c_w} \,,
  \quad
  \lambda_t  =  \frac{e}{2 s_w}\frac{m_t}{M_W} \,,
  \quad
  g_Z = \frac{e}{2 s_w c_w} \,,
  \quad 
  \alpha = \frac{e^2}{4\pi}\,.
\label{const1}
\end{eqnarray}
Here, $Q_f$ and $T_3^f$ are the fermion charge and weak isospin,
$e$ is the electric charge and $s_w$ ($c_w$) the sine (cosine) of the
Weinberg angle. 

The functions $c_i(\nu)$ are the hard QCD Wilson coefficients and depend on
$m_t, m_H$ and the c.m.\,energy $\sqrt{s}$. They also depend on the
renormalization parameter $\nu$ which accounts for the
renormalization group running of the effective currents that produce and
annihilate the $t\bar t$ 
pair in the various S-wave spin states. To achieve reliable predictions the
renormalization scaling parameter $\nu$ has to be chosen of order $\alpha_s$,
i.e.\,of order of the average top velocity in the $t\bar t$ c.m.\,system. For
this choice the imaginary part of the
zero-distance Green function does not contain any large logarithms
from ratios of the hard scales and the small nonrelativistic scales, the top
three momentum $\bmp_t\sim m_t v$ and the top kinetic energy $E_t\sim m_t v^2$
defined in the $t\bar t$ c.m.\,system. All large logarithms are summed into
the hard QCD coefficients. At NLL order the renormalization group
evolution of the hard QCD coefficients can be parameterized as
\begin{eqnarray}
c_{(1,i),\pm}(\nu) & = & c_{(1,i),\pm}(1)\,\exp\left(f(\nu,2) \right) \,,
\qquad (i=0,\pm 1)
\nonumber\\[4mm]
c_{0}(\nu) & = & c_{0}(1)\,\exp\left(f(\nu,0) \right)\,.
\label{currentWilson}
\end{eqnarray}
The function $f$ was given in Ref.~\cite{FarrellHoang1} using the results
obtained in Refs.~\cite{HoangStewartultra,Pineda1}. Whereas the
renormalization group running of the coefficients can be determined within the
effective theory, and is independent 
of the short distance process, the matching conditions at $\nu=1$ are
process-dependent. We use the convention that the LL
matching conditions for the  $c_i(\nu)$ are normalized to unity. At NLL order
the matching 
conditions are obtained from matching the factorization formula expanded to
order $\alpha_s$ to the corresponding full theory Higgs energy
distribution at ${\cal O}(\alpha_s)$ in the endpoint region expanded to
${\cal O}(v)$ for stable top quarks and using $\nu=1$ ($\mu=m_t$) for the
renormalization scaling parameters. The full theory predictions are taken from the
numerical codes obtained in Ref.~\cite{Denner1}. More 
information on the numerical matching procedure can be found in
Ref.~\cite{FarrellHoang1}. The NLL matching conditions can be parameterized in
the form 
\begin{eqnarray}
c_{(1,i),\pm}(\nu=1) & = & 1 + \frac{C_F\alpha_s(m_t)}{2}\,\delta
c_{(1,i),\pm}(\sqrt{s},m_t,m_H) \,,
\qquad (i=0,\pm 1)
\nonumber\\[2mm]
c_{1,\pm}(\nu=1) & = & 1 + \frac{C_F\alpha_s(m_t)}{2}\,\delta
c_{1,\pm}(\sqrt{s},m_t,m_H) \,,
\nonumber\\[2mm]
c_{0}(\nu=1) & = & 1 + \frac{C_F\alpha_s(m_t)}{2}\,\delta
c_{0}(\sqrt{s},m_t,m_H)
\,,
\label{matchcond}
\end{eqnarray}
and numerical results for the NLL order contributions for various choices
of $\sqrt{s}$, $m_t$ and $m_H$ are given in Tab.~\ref{tab1}.
\tabcolsep1.5mm
\begin{table}
\begin{center} 
\begin{tabular}{|c||c|c||l|l||l|l|l|l|l|}\hline   
$\sqrt s$ & $m_t$ & $m_H$ & \multicolumn{1}{|c|}{$\delta c_{1,+}$} &  
\multicolumn{1}{|c|}{$\delta c_{1,-}$} & \multicolumn{1}{|c|}{$\delta  
c_{(1,\pm1),+}$} & \multicolumn{1}{|c|}{$\delta c_{(1,\pm1),-}$} &  
\multicolumn{1}{|c|}{$\delta c_{(1,0),+}$} &  
\multicolumn{1}{|c|}{$\delta c_{(1,0),-}$}&  
\multicolumn{1}{|c|}{$\delta c_{0}$}\\ \hline \hline 
 500 & 170 & 115 & -2.3011(2) & -2.2703(2) & -2.2954(2) & -2.2654(2) &  
-2.3134(2) & -2.2807(2) & -0.573(4) \\ \hline 
 490 & 170 & 115 & -2.2910(4) & -2.2618(4) & -2.2867(4) & -2.2581(4) &  
-2.3001(4) & -2.2695(4) & -0.565(5) \\ \hline 
 480 & 170 & 115 & -2.2804(7) & -2.2528(7) & -2.2775(7) & -2.2503(7) &  
-2.2866(7) & -2.2581(7) & -0.557(6) \\ \hline 
 470 & 170 & 115 & -2.2689(5) & -2.2430(5) & -2.2672(5) & -2.2415(5) &  
-2.2724(5) & -2.2460(5) & -0.547(9) \\ \hline 
 460 & 170 & 115 & -2.257(1) & -2.232(1) & -2.256(1) & -2.232(1) &  
-2.258(1) & -2.233(1) & -0.54(1) \\ \hline 
 \hline 
 500 & 170 & 120 & -2.2992(4) & -2.2681(4) & -2.2940(4) & -2.2637(4) &  
-2.3105(4) & -2.2776(4) & -0.572(4) \\ \hline 
 490 & 170 & 120 & -2.2890(6) & -2.2596(6) & -2.2852(6) & -2.2563(6) &  
-2.2971(6) & -2.2664(6) & -0.564(5) \\ \hline 
 480 & 170 & 120 & -2.2779(4) & -2.2501(4) & -2.2754(4) & -2.2479(4) &  
-2.2830(4) & -2.2544(4) & -0.555(4) \\ \hline 
 470 & 170 & 120 & -2.2660(9) & -2.2399(9) & -2.2648(9) & -2.2389(9) &  
-2.2684(9) & -2.2419(9) & -0.546(9) \\ \hline 
 \hline 
 500 & 170 & 140 & -2.2931(6) & -2.2610(6) & -2.2901(6) & -2.2584(6) &  
-2.2994(6) & -2.2663(6) & -0.568(9) \\ \hline 
 490 & 170 & 140 & -2.2815(6) & -2.2510(6) & -2.2800(6) & -2.2498(6) &  
-2.2845(6) & -2.2536(6) & -0.559(9) \\ \hline 
 \hline 
 500 & 175 & 115 & -2.2871(3) & -2.2605(3) & -2.2831(3) & -2.2571(3) &  
-2.2956(3) & -2.2678(3) & -0.562(2) \\ \hline 
 490 & 175 & 115 & -2.2767(4) & -2.2516(4) & -2.2740(4) & -2.2492(4) &  
-2.2824(4) & -2.2565(4) & -0.554(2) \\ \hline 
 480 & 175 & 115 & -2.2657(6) & -2.2421(6) & -2.2641(6) & -2.2407(6) &  
-2.2689(6) & -2.2449(6) & -0.544(9) \\ \hline 
 470 & 175 & 115 & -2.2536(9) & -2.2315(9) & -2.2531(9) & -2.2311(9) &  
-2.2546(9) & -2.2324(9) & -0.54(1) \\ \hline 
 \hline 
 500 & 175 & 120 & -2.2848(5) & -2.2580(5) & -2.2813(5) & -2.2550(4) &  
-2.2923(4) & -2.2645(4) & -0.561(5) \\ \hline 
 490 & 175 & 120 & -2.2741(5) & -2.2488(5) & -2.2719(5) & -2.2469(5) &  
-2.2789(5) & -2.2529(5) & -0.553(4) \\ \hline 
 480 & 175 & 120 & -2.263(1) & -2.2389(8) & -2.2616(8) & -2.2380(8) &  
-2.265(1) & -2.2409(8) & -0.544(6) \\ \hline 
 \hline 
 500 & 175 & 140 & -2.2766(5) & -2.2489(5) & -2.2752(5) & -2.2477(5) &  
-2.2793(5) & -2.2512(5) & -0.556(5) \\ \hline 
 \end{tabular} 
\end{center}
\caption{Numerical values for the matching conditions 
for the singlet and triplet hard QCD coefficients for typical values
$\sqrt{s}$, $m_t$ and $m_H$. The masses and energies are given in units of
GeV. Note that $c_{(1,+1),\pm}=c_{(1,-1),\pm}$ due to parity.}
\label{tab1}
\end{table}
The singlet matching conditions do not depend on the 
electron-positron polarization  because there is only one non-trivial
QCD form factor in the full theory 
that can contribute to the hard QCD matching conditions for the
effective theory spin singlet $t\bar t$ current. In Feynman gauge it
originates from the pseudoscalar Goldstone-$t\bar t$ vertex.  
For the triplet currents, on the other hand, several form 
factors contribute in the full theory $t\bar t$ vertices, therefore  the  
matching conditions are polarization-dependent for the parameterization used in
Eq.~(\ref{dsdEHEFT}). 

If the polarization of the $t\bar t$ final states is not
accounted for, the factorization formula can be written in a simpler form
using for the $t\bar t$ spin triplet contributions the definitions
\begin{eqnarray}
c_{1,\pm}^2(\nu)\,F_{1,\pm}^{\gamma Z} 
& \equiv & 
\sum_{i=-1,0,+1}c_{(1,i),\pm}^2(\nu)\,F_{(1,i),\pm}^{\gamma Z}
\,, \qquad
F_{1,\pm}^{\gamma Z} \, \equiv \,
\sum_{i=-1,0,+1}\,F_{(1,i),\pm}^{\gamma Z}
\,,
\nonumber\\[2mm]
c_{1,\pm}^2(\nu) & = &
\frac{\sum_{i=-1,0,+1}c_{(1,i),\pm}^2(\nu)\,F_{(1,i),\pm}^{\gamma Z}}
{F_{1,\pm}^{\gamma Z}}
\,.
\label{averagedef}
\end{eqnarray}
In Ref.~\cite{FarrellHoang1} the results for the triplet contributions were
presented in this form.

\section{The Low Higgs Energy Endpoint Region} 
\label{sectionlowHE}

\begin{figure}[t] 
  \begin{center}
    \epsfig{file=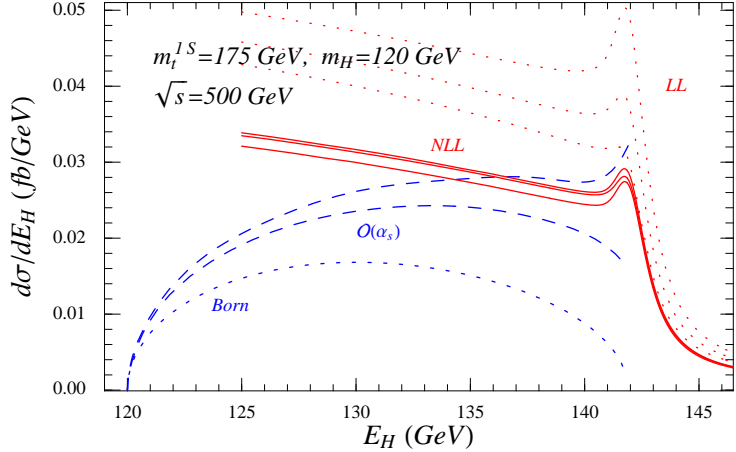,height=6cm}
    \caption{The unpolarized Higgs energy spectrum in the
    nonrelativistic expansion at LL (dotted lines) and NLL (solid
    lines) order for $\nu=0.1,\,0.2,\,0.4$. The fixed-order expansion
    is also shown at Born level (lower dotted line) and at $\mathcal
    O(\alpha_s)$ for $\mu=\sqrt s$ (lower dashed line) at for
    $\mu=\sqrt s\,v$ (upper dashed line).  The cross section at NLL
    order fails to reproduce the correct physical behavior of the 
    fixed-order results from the loop expansion in the {\it low} Higgs
    energy regime. At the 1S peak the upper (lower) NLL order 
    curve corresponds to the effective theory renormalization 
    parameter $\nu=0.2(0.1)$.    
      \label{fig2} }
  \end{center}
\end{figure}

In Fig.~\ref{fig2} the prediction for the unpolarized Higgs energy spectrum
obtained from the factorization formula in Eq.~(\ref{dsdEHEFT}) has been
displayed at LL (dotted lines) and NLL (solid lines) order in the
nonrelativistic expansion for the effective theory renormalization parameters
$\nu=0.1,0.2,0.4$. The parameters are $\sqrt{s}=500$~GeV, $m_t^{\rm
  1S}=175$~GeV, $m_H=120$~GeV, and  
\begin{eqnarray}
\begin{array}{ll}
\Gamma_t=1.43~\mbox{GeV}\,, &  \\
M_Z=91.1876~\mbox{GeV}\,,   & \quad M_W=80.423~\mbox{GeV}, \\
\alpha^{-1}=137.036\,,      & \quad c_w=M_W/M_Z\,.
\end{array}
\label{parameters}
\end{eqnarray}
%
%
We have also plotted the tree level (lower dotted line) and the 
${\cal O}(\alpha_s)$ Higgs energy spectrum for $\mu=\sqrt{s}$ (lower dashed
line) and for $\mu=\sqrt s\, v$ (upper dashed line) where $v$ is the
$t\bar t$ relative velocity defined in Eq.~\eqref{vdef}. Since the hard scale
as well as the relative momentum of the top quarks are scales that are relevant 
for nonrelativistic $t\bar t$ production, the difference between the two scale
choices illustrates the ambiguity contained in the fixed-order calculation
close to the large Higgs energy
endpoint. A detailed discussion of the deficiencies of the fixed-order
predictions in the endpoint region and quality of the nonrelativistic
expansion and the theoretical normalization uncertainty of the NLL
order prediction has been given in Ref.~\cite{FarrellHoang1} and shall
not be repeated here. The issue 
we want to point out in Fig.~\ref{fig2} is that the predictions obtained
from the factorization formula in Eq.~(\ref{dsdEHEFT}), which properly
accounts for the summation of all NLL order contributions in the 
{\it large} Higgs energy region,
is not compatible with the correct physical behavior at the {\it low} Higgs
energy endpoint $E_H=m_H$. There the Higgs boson is produced at rest (in the
lab frame) and the Higgs energy spectrum has to go to zero as do
the tree level and ${\cal O}(\alpha_s)$ predictions. In particular, at
the low Higgs energy endpoint region there is no singular enhancement
from the matrix elements, and due to phase space suppression the
coefficient functions $G_i$ of e.g. the tree level Higgs energy spectrum
(see Appendix \ref{app1}) vanish like   
$G_i \sim \hat\beta$ with
\begin{equation}
\hat\beta \, = \,
\left(\,
\frac{ m_H\,(\sqrt{s} - m_H)^2\,(\,(\sqrt{s} - m_H)^2 - 4m_t^2\,)}
{m_t^2\, s^{3/2}}
\,\right)^{1/2}\,\sqrt{ v_{\rm max}^2-v^2 } 
\, + \, {\cal O}( v_{\rm max}^2-v^2)^{3/2}
\,.
\end{equation}
This endpoint behavior cannot be obtained within the nonrelativistic expansion
in small $v$ even if the endpoint is located at a velocity much smaller than
one, see Eq.~(\ref{vmaxdef}). Terms that are formally from beyond NLL order
in $v$ thus need to be summed up to achieve a correct low Higgs energy
endpoint behavior. 

It is useful for the construction of a factorization formula which
can account for the correct physical low Higgs energy behavior that the
full theory 
tree level Higgs energy spectrum, both for the $t\bar t$ pair in the spin
singlet and for the (combined) triplet configuration, does not have
order $v$ (NLL) corrections to the  
leading endpoint behavior in the large Higgs energy endpoint, i.e.
\begin{eqnarray}
\left(\frac{d\sigma}{d E_H}(E_H\approx E_H^{0})\right)^\pm_{1, {\rm Born}} & = &
\left[\frac{2\,N_c\,\left[(1+x_H-4x_t)^2-4x_H\right]^{1/2}}{s^{3/2}\,\pi}\,
F_{1,\pm}^{\gamma Z}\,\right] v \, + \, {\cal O}(v^3)
\,,
\nonumber \\[4mm]
\left(\frac{d\sigma}{d E_H}(E_H\approx E_H^{0})\right)^\pm_{0, {\rm Born}} & = &
\left[\frac{2\,N_c\,\left[(1+x_H-4x_t)^2-4x_H\right]^{1/2}}{s^{3/2}\,\pi}\,
F^Z_{0,\pm}\,\right] v \, + \, {\cal O}(v^3)
\,.
\end{eqnarray}
At NLL order it is thus consistent to use the full tree level $E_H$ spectrum
in the large Higgs energy endpoint 
instead of the constant LL matching conditions $F^{Z,\gamma Z}$
given in Eqs.~(\ref{F1def})-(\ref{F0def}),
\begin{eqnarray}
F_{1,\pm}^{\gamma Z} & \to &
\left(\frac{d\sigma}{d E_H}\right)^\pm_{\rm Born}\,
\frac{F_{1,\pm}^{\gamma Z}}{F_{0,\pm}^{Z}+F_{1,\pm}^{\gamma Z}}\,
\left[\frac{2\,N_c\,\left[(1+x_H-4x_t)^2-4x_H\right]^{1/2}}{s^{3/2}\,\pi}\,v
\,\right]^{-1}
\,,
\nonumber\\[4mm]
F_{0,\pm}^{Z} & \to &
\left(\frac{d\sigma}{d E_H}\right)^\pm_{\rm Born}\,
\frac{F_{0,\pm}^{Z}}{F_{0,\pm}^{Z}+F_{1,\pm}^{\gamma Z}}\,
\left[\frac{2\,N_c\,\left[(1+x_H-4x_t)^2-4x_H\right]^{1/2}}{s^{3/2}\,\pi}\,v
\,\right]^{-1}
\,,
\label{modifiedrules}
\end{eqnarray} 
where $(\frac{d\sigma}{d E_H})^\pm_{\rm Born}$ is the full tree level 
Higgs energy spectrum for the polarized $e^+e^-$ initial state. Note that the
replacement prescription in Eq.~(\ref{modifiedrules}) can only be applied for
Higgs energies smaller than $E_H^0$, for larger Higgs energies
Eq.~\eqref{dsdEHEFT} is left unchanged. 
For the convenience of the reader we have given the analytic expressions for
the tree level Higgs energy spectrum in the appendix using up to minor
modifications the conventions of Ref.~\cite{Dawson2}. They also correct a
few typos that were contained in Ref.~\cite{Dawson2} and pointed out
before in Ref.~\cite{FarrellHoang1}. For the case of an unpolarized
$t\bar t$ final state, using the
prescription~(\ref{modifiedrules}) in the factorization
formula~(\ref{dsdEHEFT}) leads to a modified factorization formula that
resums correctly all NLL order terms. In addition it has 
the correct physical behavior at the low Higgs energy endpoint
$E_H=m_H$. 

The modified NLL factorization formula based on Eqs.~(\ref{dsdEHEFT}),
(\ref{averagedef}), and (\ref{modifiedrules}) is not unique, 
alternative prescriptions to achieve the correct physical low Higgs energy
endpoint behavior are conceivable. However, different prescriptions will only
affect the low Higgs energy endpoint where the $E_H$ spectrum vanishes, and
they should therefore not have a large numerical impact. While the modified
NLL factorization formula contains the exact tree level contribution, its 
${\cal O}(\alpha_s)$ 
contribution (in the expansion in powers of $\alpha_s$) differs from the exact
${\cal O}(\alpha_s)$ result obtained in Ref.~\cite{Denner1} since it includes
only the QCD corrections of the large Higgs energy endpoint. Thus an estimate
of the intrinsic uncertainty in our prescription can be gained by comparing
its ${\cal O}(\alpha_s)$ terms  with the exact result from
Refs.~\cite{Denner1,Dittmaier1}. For stable and unpolarized top quarks
the first two terms in the $\alpha_s$ expansion of our modified
factorization formula read 
\begin{eqnarray}
\left(\frac{d\sigma}{d E_H}(E_H)\right)^\pm_{\rm NLL} & = &
\left(\frac{d\sigma}{d E_H}(E_H)\right)^\pm_{\rm Born} + 
\left(\frac{d\sigma}{d E_H}(E_H)\right)^\pm_{{\cal O}(\alpha_s)} +
{\cal O}(\alpha_s^2)
\,,
\label{Oasexpand}
\end{eqnarray}
where 
\begin{eqnarray}
\left(\frac{d\sigma}{d E_H}(E_H)\right)^\pm_{{\cal O}(\alpha_s)} & = &
C_F\alpha_s\,
\bigg[\,
  \frac{F_{0,\pm}^{Z}\delta c_0+F_{1,\pm}^{\gamma Z}\delta c_{1,\pm}}
    {F_{0,\pm}^{Z}+F_{1,\pm}^{\gamma Z}}
 + \frac{\pi}{2}\bigg(1-\frac{4m_t^2}{Q^2}\bigg)^{-1/2}
\,\bigg]
\,\left(\frac{d\sigma}{d E_H}(E_H)\right)^\pm_{\rm Born}
\,.
\nonumber\\ &&
\label{Oasapprox}
\end{eqnarray}

In Tab.~\ref{tab2} numerical results for the exact total ${\cal O}(\alpha_s)$
unpolarized cross section, $\sigma_{\rm exact}^{{\cal O}(\alpha_s)}$~\cite{Denner1}, 
and for the ${\cal O}(\alpha_s)$ approximation
from Eq.~(\ref{Oasexpand}), $\sigma_{\rm NLL}^{{\cal O}(\alpha_s)}$, 
are shown for various c.m.\,energies and 
$m_t=175$~GeV, $m_H=120$~GeV, $\Gamma_t=0$, $\mu=\sqrt{s}$. 
For c.m.\,energies below $500$~GeV the 
deviation increases with the c.m.\,energy. It vanishes at the
three-body threshold $\sqrt{s}\approx 2m_t+m_H$ and reaches the level
of 1.5\% for $\sqrt{s}=500$~GeV.  

\begin{table}
\begin{center}
\begin{tabular}{|c||c|c|c|}\hline  $\sqrt s$ &  
$\sigma_{\rm exact}^{\alpha_s}$ & $\sigma_{\rm NLL}^{\alpha_s}$ &  
rel. dev. $(\%)$ \\ \hline \hline 
 475 &  0.0311 &  0.0309 &  0.6 \\ \hline 
 480 &  0.0908 &  0.0900 &  0.9 \\ \hline 
 490 &  0.254 &  0.251 &  1.2 \\ \hline 
 500 &  0.446 &  0.439 &  1.5 \\ \hline 
 550 &  1.366 &  1.343 &  1.7 \\ \hline 
 600 &  1.953 &  1.924 &  1.5 \\ \hline 
 700 &  2.356 &  2.348 &  0.4 \\ \hline 
 \end{tabular} 
\end{center}
\caption{The total cross section using the exact $\mathcal
  O(\alpha_s)$  result $\sigma_{\rm exact}^{\mathcal
  O(\alpha_s)}$ from Ref.~\cite{Denner1} and  the
  approximation based on Eq.~(\ref{Oasexpand}), $\sigma_{\rm NLL}^{\mathcal
  O(\alpha_s)}$. The third column
  shows the relative deviation in percent. 
  The difference between the two calculations is
  maximal for c.m.\,energies around $550$~GeV. }
\label{tab2} 
\end{table}

\begin{figure}[t] 
  \begin{center}
    \epsfig{file=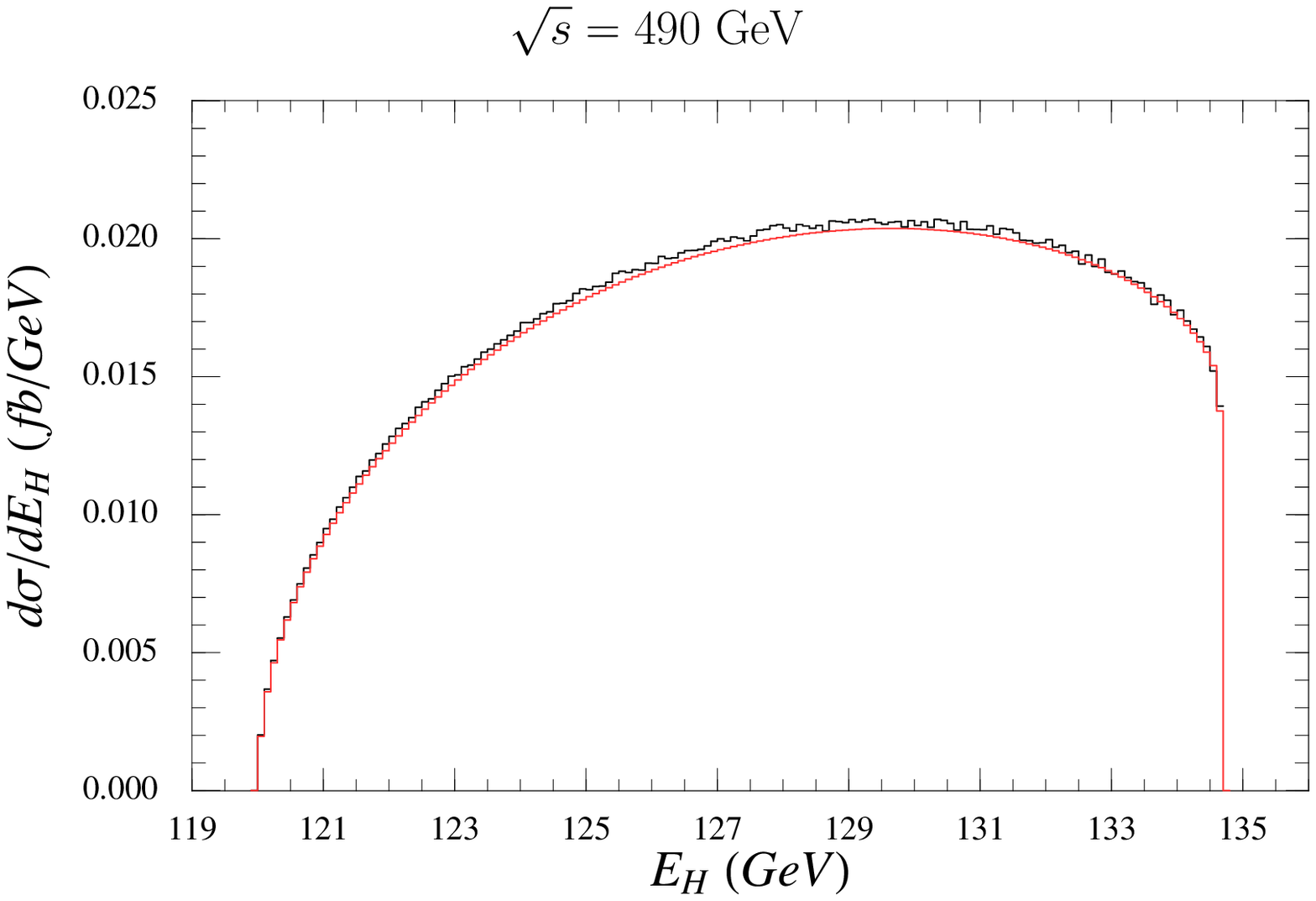,height=4.5cm}\hspace{0.5cm}
    \epsfig{file=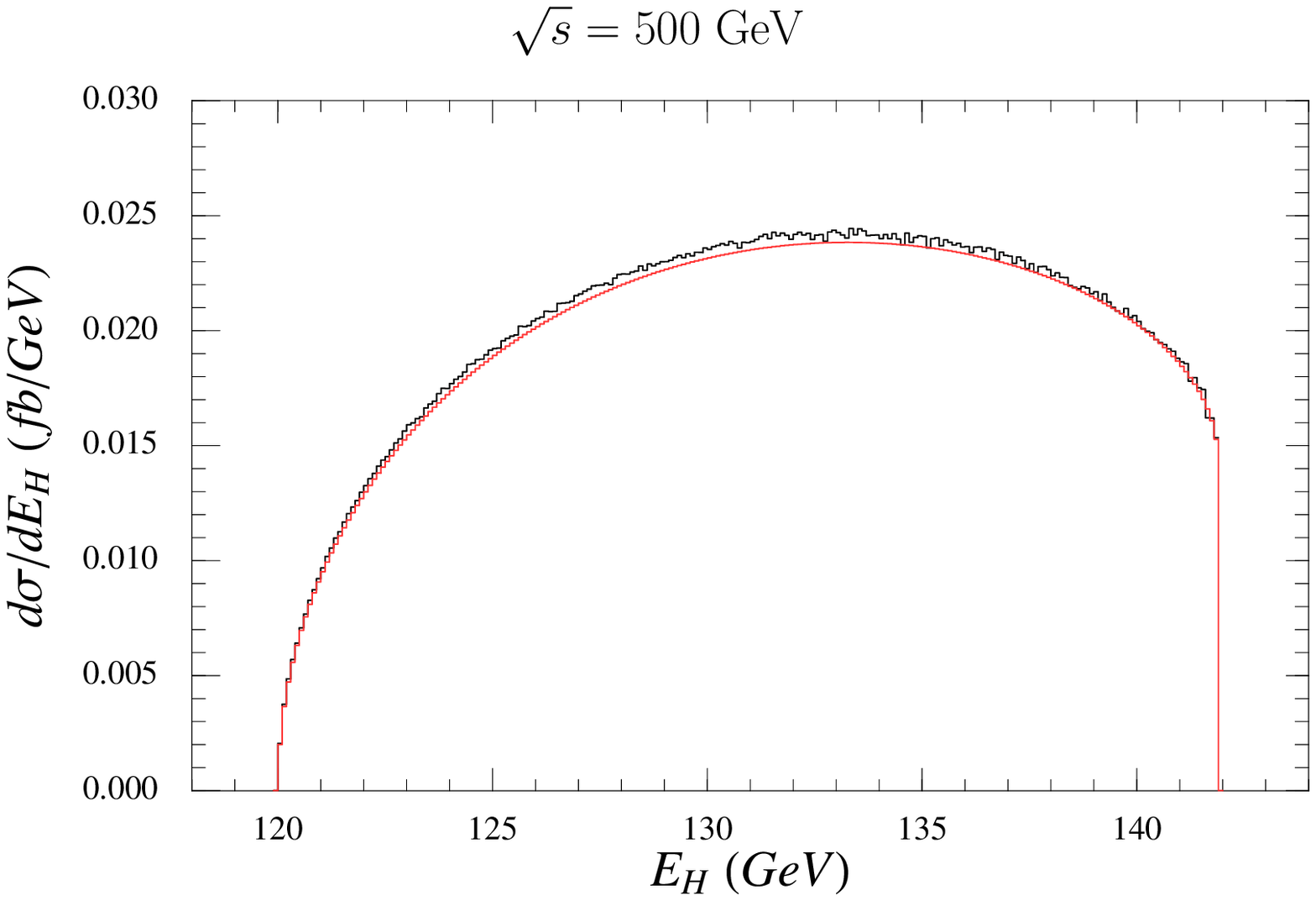,height=4.5cm}\\[1em]
    \epsfig{file=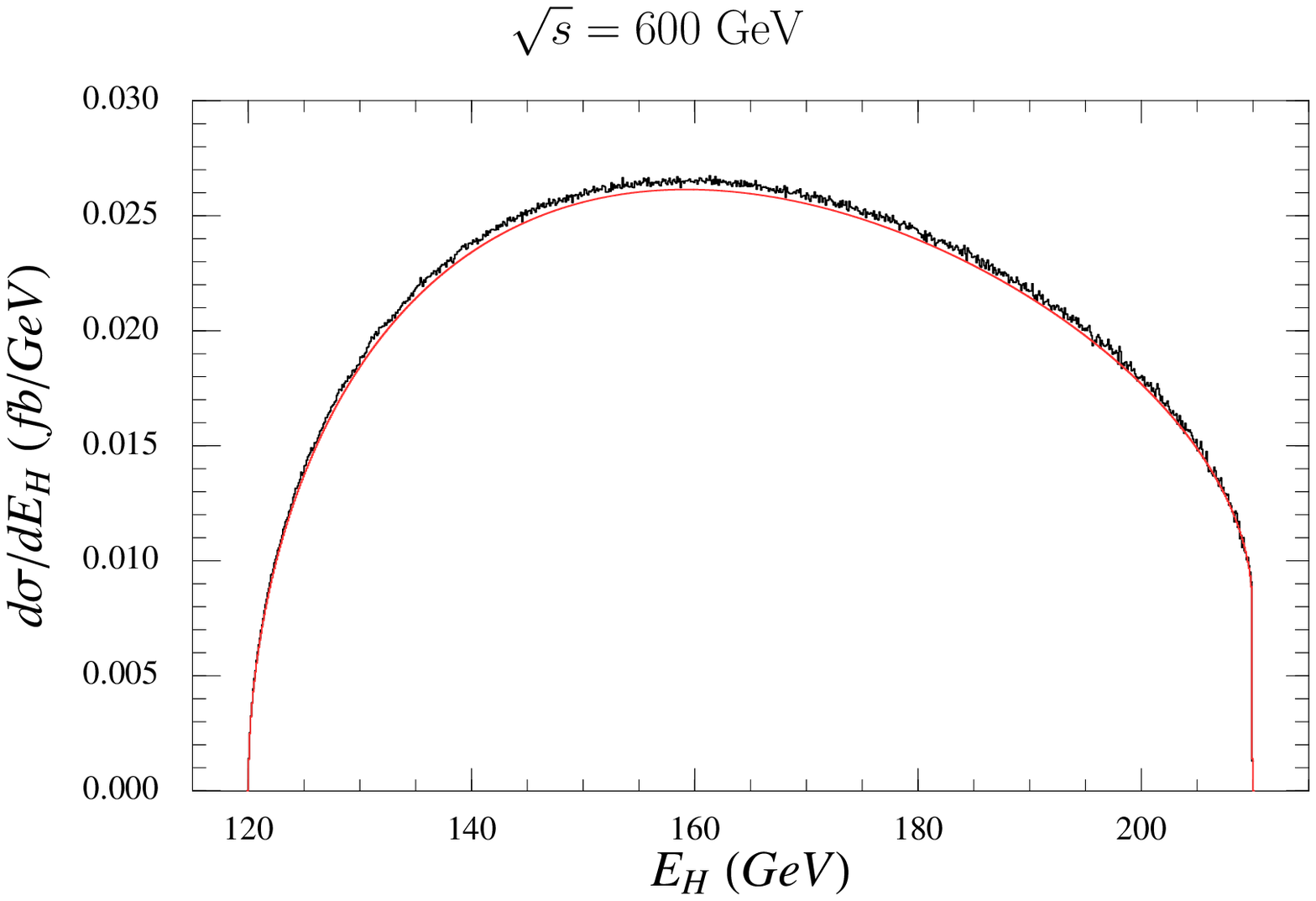,height=4.5cm}\hspace{0.5cm}
    \epsfig{file=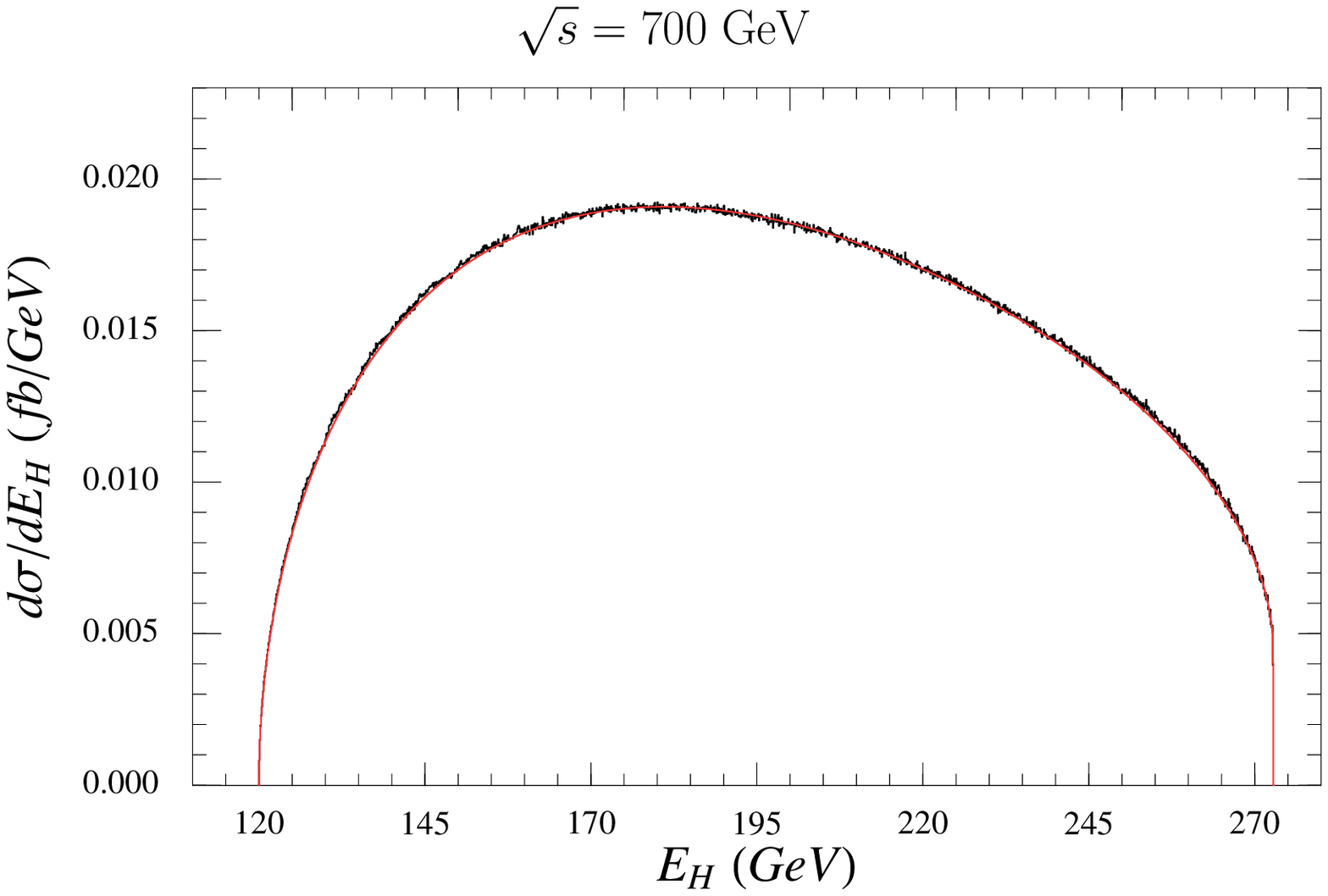,height=4.5cm}
    \caption{
The exact ${\cal O}(\alpha_s)$ unpolarized Higgs energy
spectrum from Ref.~\cite{Denner1} (black lines) and the 
${\cal O}(\alpha_s)$ approximation in  Eq.~(\ref{Oasexpand}) 
(gray lines) for different c.m.\,energies $\sqrt s$ for $m_t=175$~GeV,
$m_H=120$~GeV, and $\mu=\sqrt s$. 
      \label{fig3} }
  \end{center}
\end{figure}

In Fig.~\ref{fig3} the exact ${\cal O}(\alpha_s)$ unpolarized Higgs energy
spectrum (black lines) and the ${\cal O}(\alpha_s)$ approximation in
Eq.~(\ref{Oasexpand}) (gray lines) are displayed in $0.1$~GeV bins for
$\sqrt{s}=490, 500, 600$, and $700$~GeV,  $m_t=175$~GeV,
$m_H=120$~GeV, $\Gamma_t=0$, and $\mu=\sqrt s$. Note that for the strong
coupling we use $\alpha_s(500~\mbox{GeV})=0.09396$. The other 
parameters are chosen as in Eq.~(\ref{parameters}). 
For $\sqrt s = 500$ GeV the relative deviation in the Higgs energy spectrum is
at most $2.8\%$. The difference is smaller for lower c.m.\,energies since the
maximal possible top relative velocity $v^{\rm max}$ is increasing
with the c.m.\,energy, see Eq.~(\ref{vmaxdef}).  The results indicate
that the intrinsic  uncertainty of our approach is substantially
smaller than the theoretical uncertainty of $5\%$ from uncalculated
higher order QCD effects~\cite{FarrellHoang1,HoangEpi}. 
%
%

In Figs.~\ref{fig3} c and d and in Tab.~\ref{tab2} we have  analyzed the
difference between the exact ${\cal O}(\alpha_s)$ results and the 
${\cal O}(\alpha_s)$ approximation based on Eq.~(\ref{Oasexpand}) for larger
c.m.\,energies as well. It is a surprising fact that the fairly simple
expression 
in Eq.~(\ref{Oasapprox}), which contains only tree level information and
the NLL QCD information from the large Higgs energy endpoint, can
also account very well for the exact ${\cal O}(\alpha_s)$ results at higher
energies, where real gluon radiation is non-negligible. For
c.m.\,energies between $500$ and $700$~GeV the approximation based on
Eq.~(\ref{Oasexpand}) deviates from  the exact results 
by at most $1.8\%$ for the unpolarized total cross section,
where the maximal deviation is reached for $\sqrt{s}\approx
550$~GeV. Since the numerical evaluation of Eq.~(\ref{Oasapprox}) is
substantially faster than for the exact ${\cal O}(\alpha_s)$
result~\cite{Denner1}, it can be useful as an efficient approximation
formula for higher c.m.\,energies.

\section{Numerical Analysis} 
\label{sectionanalysis}
%
%

%
%
\begin{figure}[t] 
  \begin{center}
    \epsfig{file=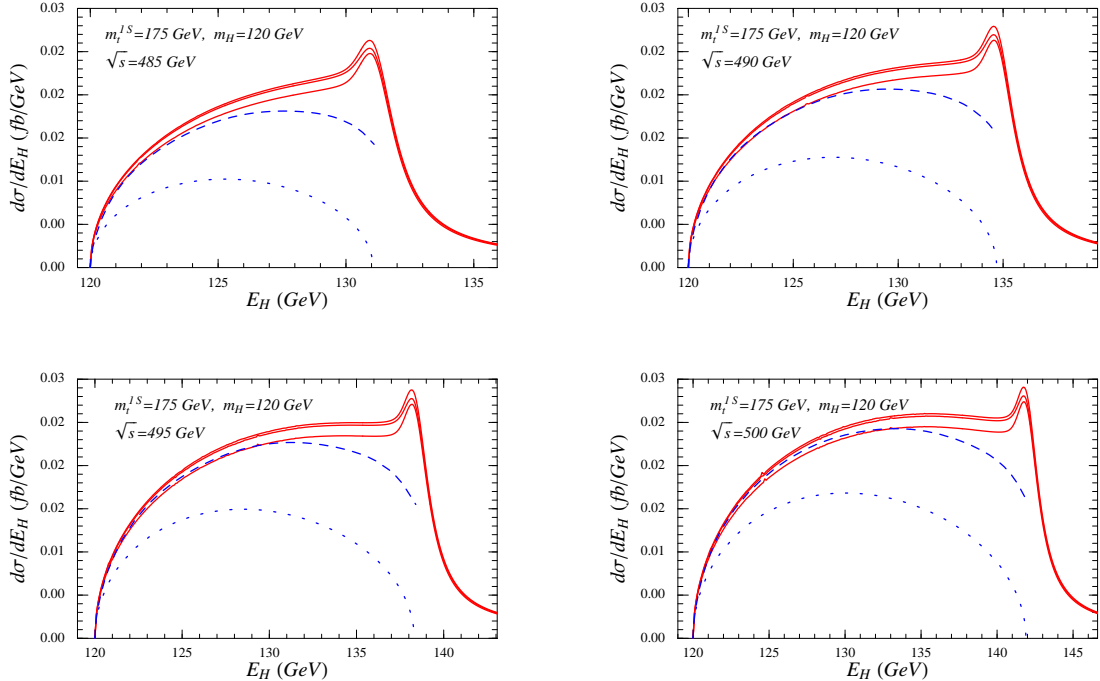,height=9.5cm}
    \caption{The unpolarized Higgs energy spectrum  for different 
      c.m.\,energies at NLL order (solid
      lines) using the modified factorization formula 
      based on Eqs.~(\ref{dsdEHEFT}), (\ref{averagedef}), and 
      (\ref{modifiedrules}) for the renormalization
      parameters $\nu=0.1,0.2,0.4$, at ${\cal O}(\alpha_s)$ (dashed
      lines) from Ref.~\cite{Dittmaier1} with $\mu=\sqrt s$, and at
      Born level (dotted line). At the 1S peak the upper (lower) NLL 
      order curve corresponds to the effective theory renormalization 
      parameter $\nu=0.2(0.1)$.
      \label{fig4} }
  \end{center}
\end{figure}

In Fig.~\ref{fig4} the unpolarized Higgs energy spectrum at NLL order (solid
lines) using the modified factorization formula 
based on Eqs.~(\ref{dsdEHEFT}), (\ref{averagedef}), and (\ref{modifiedrules})
is displayed for the renormalization parameters $\nu=0.1,0.2,0.4$ for
the c.m.\,energies $\sqrt{s}=485,490,495,500$~GeV and 
$m_t=m_t^{\rm 1S}=175$~GeV, $m_H=120$~GeV.  The other parameters are chosen
as in Eq.~(\ref{parameters}). For comparison we also show the tree level
prediction (dotted lines) and the ${\cal O}(\alpha_s)$
results~\cite{Denner1} (dashed lines) with $\mu=\sqrt s$ for a stable
top quark. The nonrelativistic NLL order results show a substantial
enhancement  compared to the tree level and one-loop QCD predictions. 
The Higgs energy spectrum in the effective theory extends beyond the endpoint
$E_H^0$ that is obtained for the stable top quark case.
This is because the top quarks can be produced off-shell with
invariant masses smaller than $m_t$ if the top quark decay is accounted
for. With the present technology the finite top quark lifetime can only be
implemented systematically in an expansion in the top quark
off-shellness, which is naturally provided by the nonrelativistic
expansion we use here.  

It is conspicuous that the spectrum above the endpoint $E_H^0$ in the
NLL prediction falls off quite slowly. Since the average c.m.\,top
quark velocity increases with the Higgs energy for $E_H>E_H^0$ we
define the  total cross section by applying a cut on the Higgs energy above
$E_H^0$ such that the average c.m.\,top velocity remains below 
$v_{\rm cut}=0.2$. We 
fix the relation between the maximal Higgs energy and $v_{\rm cut}$ by
the relation 
$E_H^{\rm cut} = (s+m_H^2-Q^2_{\rm cut})/(2\sqrt{s})$, which is
exact in the stable top case. Here,
$Q^2_{\rm cut} \equiv (4m_t^2)/(1+v_{\rm cut}^2)$
is the minimal $t\bar t$ invariant mass. 
Note that $Q_{\rm cut}$ is smaller than $2 m_t$ because for
$E_H>E_H^0$ we are in the bound state regime. As mentioned before,
we plan a systematic treatment of finite lifetime and off-shell
effects at the NLL order level in a subsequent publication. 

In Tab.~\ref{tab3} the impact of the NLL order summations on the total cross
section for unpolarized $t\bar t$ pairs and polarized electron-positron beams
is analyzed numerically for various c.m.\,energies, top quark masses and Higgs
masses. The other parameters are chosen as in Eq.~(\ref{parameters})
except for the case $m_t=170$~GeV where we use $\Gamma_t=1.31$~GeV. 
\tabcolsep2mm
\begin{table}
\begin{center}
\begin{tabular}{|c|c|c||c|l|c||c|l|c|}\hline  $\sqrt  
s$ & $m_t$ & $m_H$ & $\sigma_{\rm Born}^+ \mbox{(fb)}$ &  
\multicolumn{1}{|c|}{$\sigma_{\rm NLL}^+ \mbox{(fb)}$} & $\sigma_{\rm  
NLL}^+/\sigma_{\rm Born}^+$ & $\sigma_{\rm Born}^- \mbox{(fb)}$&  
\multicolumn{1}{|c|}{$\sigma_{\rm NLL}^- \mbox{(fb)}$} & $\sigma_{\rm  
NLL}^-/\sigma_{\rm Born}^-$\\ \hline \hline 
$500$ & $170$ & $115$ & $ 0.644$ & $ 0.989(49)$ & $ 1.54$ & $ 1.660$ & $ 2.568(128)$ & $ 1.55$\\\hline
$490$ & $170$ & $115$ & $ 0.444$ & $ 0.754(37)$ & $ 1.70$ & $ 1.149$ & $ 1.965(98)$ & $ 1.71$\\\hline
$480$ & $170$ & $115$ & $ 0.260$ & $ 0.516(25)$ & $ 1.98$ & $ 0.674$ & $ 1.347(67)$ & $ 2.00$\\\hline
$470$ & $170$ & $115$ & $ 0.108$ & $ 0.285(14)$ & $ 2.64$ & $ 0.281$ & $ 0.747(37)$ & $ 2.66$\\\hline
$460$ & $170$ & $115$ & $ 0.014$ & $ 0.086(4)$ & $ 6.17$ & $ 0.036$ & $ 0.226(11)$ & $ 6.21$\\\hline
$500$ & $170$ & $120$ & $ 0.486$ & $ 0.783(39)$ & $ 1.61$ & $ 1.258$ & $ 2.040(101)$ & $ 1.62$\\\hline
$490$ & $170$ & $120$ & $ 0.312$ & $ 0.568(28)$ & $ 1.82$ & $ 0.809$ & $ 1.483(74)$ & $ 1.83$\\\hline
$480$ & $170$ & $120$ & $ 0.159$ & $ 0.355(17)$ & $ 2.23$ & $ 0.413$ & $ 0.929(46)$ & $ 2.25$\\\hline
$470$ & $170$ & $120$ & $ 0.046$ & $ 0.159(7)$ & $ 3.48$ & $ 0.120$ & $ 0.418(20)$ & $ 3.50$\\\hline
$500$ & $170$ & $140$ & $ 0.102$ & $ 0.229(11)$ & $ 2.24$ & $ 0.268$ & $ 0.604(30)$ & $ 2.26$\\\hline
$490$ & $170$ & $140$ & $ 0.029$ & $ 0.101(5)$ & $ 3.48$ & $ 0.076$ & $ 0.268(13)$ & $ 3.51$\\\hline
$500$ & $175$ & $115$ & $ 0.459$ & $ 0.787(39)$ & $ 1.72$ & $ 1.181$ & $ 2.039(101)$ & $ 1.73$\\\hline
$490$ & $175$ & $115$ & $ 0.268$ & $ 0.538(26)$ & $ 2.01$ & $ 0.692$ & $ 1.399(69)$ & $ 2.02$\\\hline
$480$ & $175$ & $115$ & $ 0.111$ & $ 0.298(14)$ & $ 2.68$ & $ 0.288$ & $ 0.777(38)$ & $ 2.70$\\\hline
$470$ & $175$ & $115$ & $ 0.014$ & $ 0.091(4)$ & $ 6.32$ & $ 0.037$ & $ 0.236(11)$ & $ 6.35$\\\hline
$500$ & $175$ & $120$ & $ 0.322$ & $ 0.593(29)$ & $ 1.84$ & $ 0.832$ & $ 1.541(77)$ & $ 1.85$\\\hline
$490$ & $175$ & $120$ & $ 0.164$ & $ 0.371(18)$ & $ 2.26$ & $ 0.425$ & $ 0.967(48)$ & $ 2.28$\\\hline
$480$ & $175$ & $120$ & $ 0.047$ & $ 0.167(8)$ & $ 3.54$ & $ 0.123$ & $ 0.437(21)$ & $ 3.56$\\\hline
$500$ & $175$ & $140$ & $ 0.030$ & $ 0.107(5)$ & $ 3.55$ & $ 0.079$ & $ 0.281(14)$ & $ 3.57$\\\hline
 \end{tabular} 
\end{center}
{\caption{The total cross section in units of fb at Born level for
    stable top quarks and at NLL order for unstable top quarks 
    using $\nu = 0.2$ for fully polarized electron-positron beams. The
    index refers to the polarization of the electron beam. The masses
    and $\sqrt s$ are given in units of GeV. For $m_t=(170,175)$~GeV
    we use $\Gamma_t=(1.31,1.43)$~GeV.}
\label{tab3} }
\end{table}
In Tab.~\ref{tab3}, $\sigma_{\rm Born}$ refers to the tree level cross
section for stable top quarks
(see the appendix for explicit expressions) and $\sigma_{\rm NLL}$ to the NLL
total cross section as defined above and based on the modified factorization
formula discussed in Sec.~\ref{sectionlowHE}. The NLL order predictions were
obtained for the effective theory renormalization parameter
$\nu=0.2$. The uncertainties
given for $\sigma_{\rm NLL}$ reflect the 5\% theoretical error from higher
order QCD and relativistic corrections as discussed in
Ref.~\cite{FarrellHoang1}. The results in Tab.~\ref{tab3} demonstrate the
importance of the summation of the singular terms proportional to
$(\alpha_s/v)^n$ and $(\alpha_s\ln v)^n$ that arise in the endpoint region,
and of the off-shell effects that arise from the finite top quark lifetime.
Compared to the tree level predictions the enhancement is more
pronounced for smaller c.m.\,energies and larger top or Higgs masses.  

\begin{figure}[t] 
  \begin{center}
    \epsfig{file=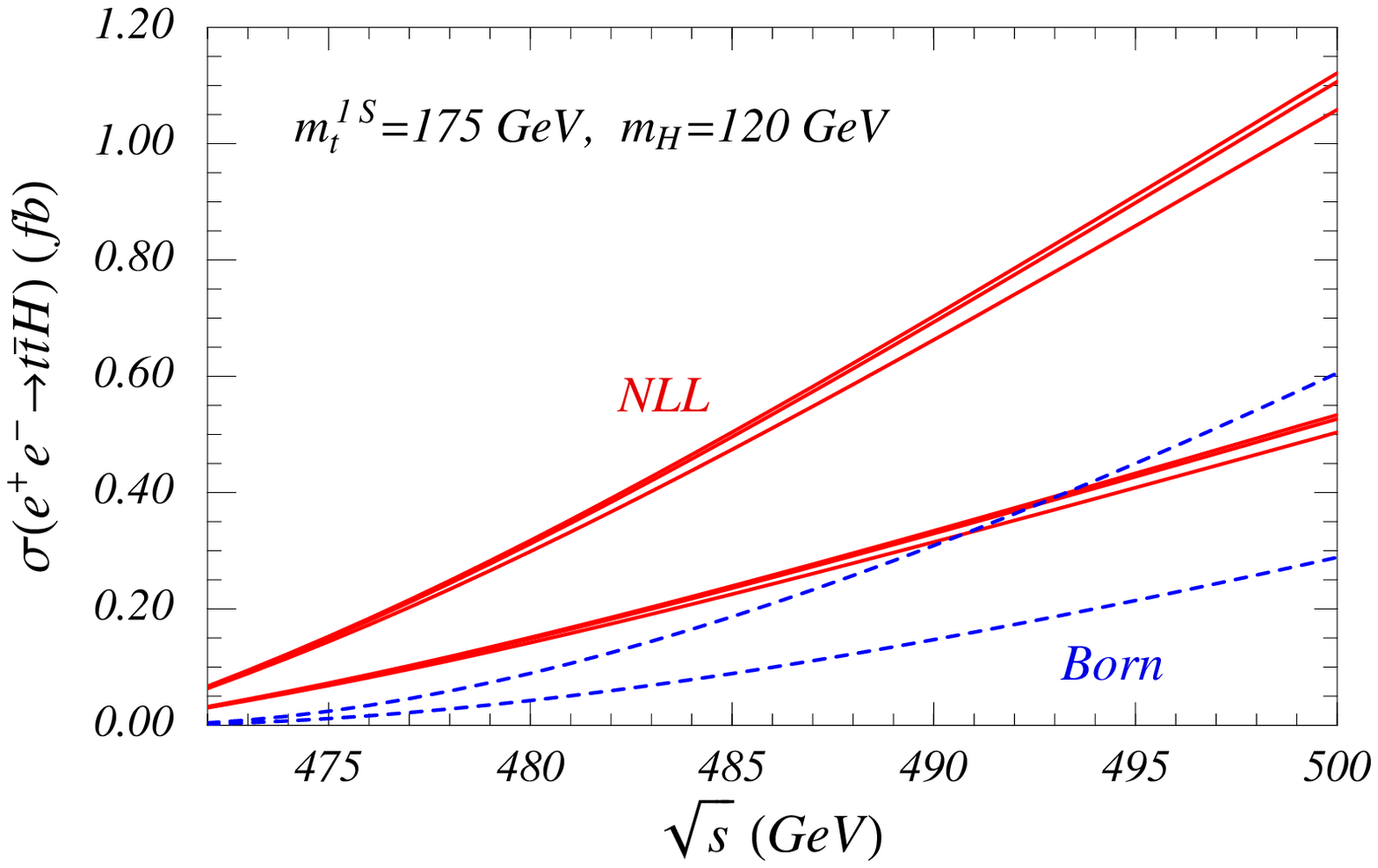,bb=80 450 560 730,height=4.7cm}
    \vspace{0.8cm}
    \epsfig{file=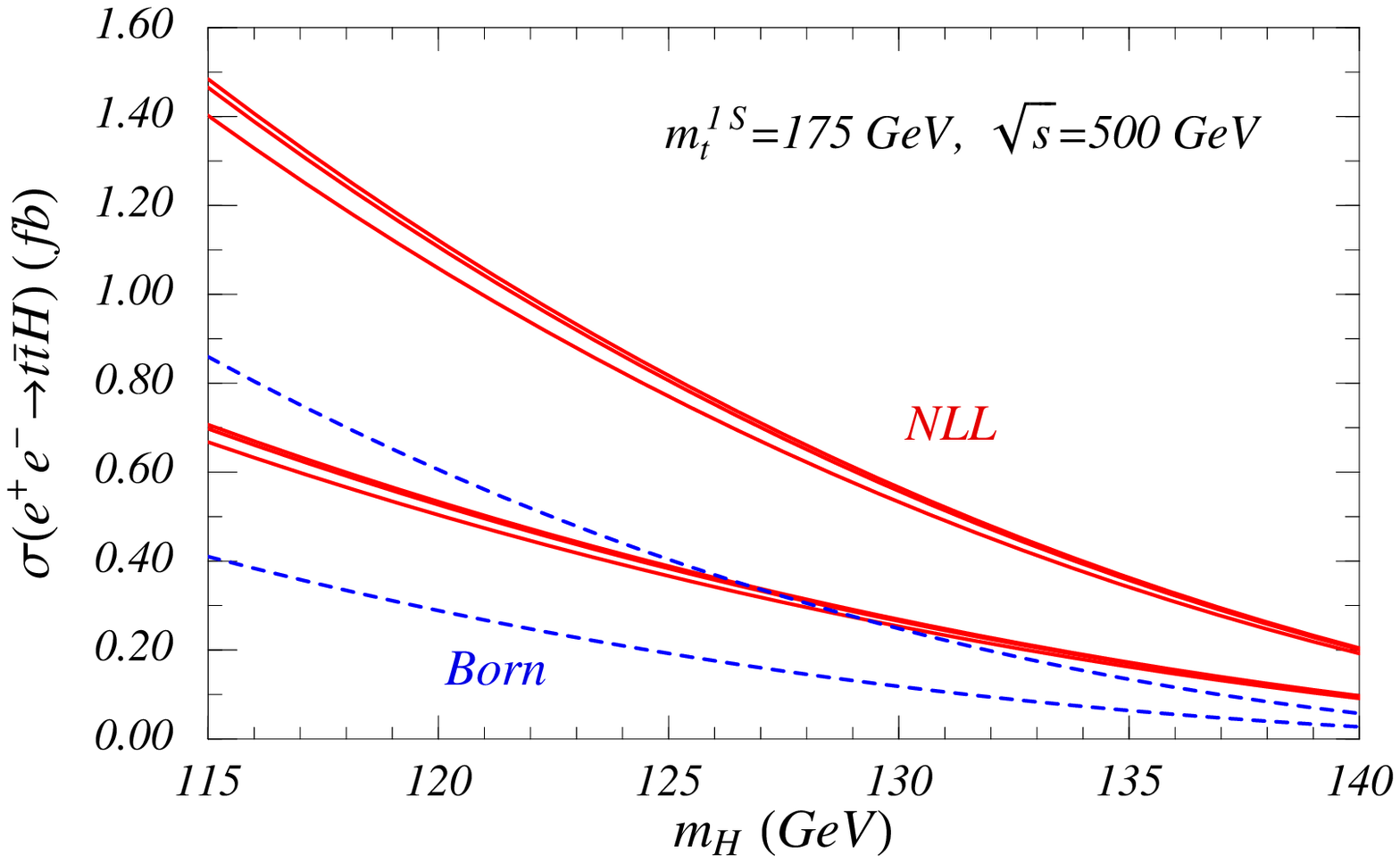,bb=80 450 560 730,height=4.7cm}\\[-2em]
    \caption{
The total cross section for unpolarized top quarks at
tree level (dashed lines) and at NLL order (solid lines) as a
function of $\sqrt{s}$ (left panel) and as a function of $m_H$ (right
panel) for unpolarized electron-positron beams $(P_+,P_-)=(0,0)$ 
(respective lower curves) and for $(P_+,P_-)=(0.6,-0.8)$ 
(respective upper curves).
      \label{fig5} }
  \end{center}
\end{figure}

It is a realistic option for the ILC project to polarize the $e^+e^-$ beams 
up to $(P_+,P_-)=(0.6,-0.8)$~\cite{TESLATDR}. Since this can
further enhance the cross section we have also assessed its merits for the
process at hand. 
In Figs.~\ref{fig5} the total cross section for unpolarized top quarks at the
tree level (dashed lines) and at NLL order (solid lines) is shown as a
function of $\sqrt{s}$ and $m_H$ for unpolarized electron-positron beams
$(P_+,P_-)=(0,0)$ and for $(P_+,P_-)=(0.6,-0.8)$. The other parameters are
chosen as in 
Eq.~(\ref{parameters}), see also the figure caption for more details. For the
NLL cross section the predictions for the three choices $\nu=0.1,0.2,0.4$ for
the renormalization scaling parameter are shown. 
%
%
The results demonstrate that using electron-positron polarization the cross
section can be enhanced by roughly a factor of two over the unpolarized cross
section. Compared to the tree level predictions 
for unpolarized electron-positron beams, which were the basis of previous
experimental analyses~\cite{Justetalk}, QCD effects and beam polarization
$(P_+,P_-)=(0.6,-0.8)$ can enhance the cross section by about a
factor of $4$ or even more for $\sqrt{s}=500$~GeV, depending on the
Higgs mass. Because of the limited statistics
for $t\bar t H$ production during the first phase of the ILC project,
these results are important for realistic experimental simulations of
Yukawa coupling measurements.  

\section{Conclusion} 
\label{sectionconclusion}

We have analyzed the impact of summing the QCD singularities proportional to
$(\alpha_s/v)^n$ and $(\alpha_s\ln v)^n$ that arise in the large Higgs energy
endpoint region for the process $e^+e^-\to t\bar t H$ for c.m.\,energies up to
$500$~GeV, i.e. energies which can be achieved during the first phase
of the ILC project. The singularities cause the breakdown of usual
multi-loop perturbation theory in powers of $\alpha_s$ and
originate from nonrelativistic dynamical QCD effects that arise because the
relative velocity of the $t\bar t$ pair is small. A consistent theoretical
treatment requires the use of nonrelativistic effective theory methods and
includes a systematic treatment of off-shell effects caused by the finite
top quark lifetime. In Ref.~\cite{FarrellHoang1} we derived a
factorization formula for the large Higgs energy endpoint region for large
c.m.\,energies above $500$~GeV. In the present work we have extended
the approach to 
c.m.\,energies below $500$~GeV, where the top quark pair is nonrelativistic in
the entire phase space, and we have also accounted for the effects of
electron-positron beam polarization. We have determined the predictions for
the Higgs energy spectrum and the total cross section at NLL order for the QCD
effects and at LL order for the top quark finite lifetime and for off-shell
effects. The QCD effects enhance the total cross section by roughly a
factor of two relative to the Born prediction for
$\sqrt{s}=500$~GeV. Using polarized electron-positron beams the cross
section can be further enhanced over the unpolarized case by another
factor of approximately two. Our results are 
important for realistic simulation studies for Yukawa coupling measurements in
the first phase of the ILC project.

\begin{acknowledgments} 
We would like to thank S.~Dittmaier and M.~Roth for 
useful discussions and for providing us their numerical codes 
from Ref.~\cite{Denner1}, and T.~Teubner for providing the TOPPIC
code. A.H. thanks A.~Juste for useful discussions and suggestions. 
\end{acknowledgments}

\vskip 1cm

\begin{appendix}
\section{Tree Level Higgs Energy Spectrum}
\label{app1}

Correcting the typos of Ref.~\cite{Dawson2} the tree level Higgs energy
spectrum in the process $e^+e^-\to t\bar t H$ for polarized electron-positron
beams reads ($x_E\equiv 2 E_H/\sqrt{s}$, 
$\sigma_{\rm pt}\equiv 4\pi \alpha^2/(3s)$)
\begin{eqnarray}
\left(\frac{d\sigma(E_H)}{d x_E}\right)_{\rm Born}^\pm & = &
\sigma_{\rm pt}\,\frac{N_c}{8\pi^2}\bigg\{\,
  \bigg[\,
   Q_e^2 Q_t^2 
   + \frac{2 Q_e Q_t (v_e \mp a_e) v_t}{1 - x_Z} 
   + \frac{(v_e \mp a_e)^2(v_t^2 + a_t^2)}{(1 - x_Z)^2}
  \,\bigg]\,G_1 
\nonumber\\[.5em] & &
  {} + \frac{(v_e \mp a_e)^2}{(1 - x_Z)^2}\, \bigg[\,
    a_t^2\sum_{i=2}^6\,G_i+v_t^2(G_4+G_6) \,\bigg]  
  + \frac{Q_e Q_t (v_e \mp a_e) v_t}{1 - x_Z}\,G_6\,\bigg\} 
\,,
\nonumber\\
\end{eqnarray}
where the coefficient functions are given by 
\begin{eqnarray}
G_1 & = & 
\frac{ 2\lambda_t^2}{(\hat\beta^2 - x_E^2)x_E}\,
\bigg\{
   -4 \hat\beta (4x_t - x_H)(2x_t + 1)x_E 
\nonumber\\[2mm] & & 
 {}+
   (\hat\beta^2 - x_E^2)\big[16 x_t^2 + 2x_H^2 - 2x_H x_E + x_E^2 
   - 4x_t(3x_H - 2 - 2x_E)\big]\,
   \ln\bigg(\frac{x_E + \hat\beta}{x_E - \hat\beta}\bigg) \bigg\}
,\qquad
\\[4mm]
G_2 & = &
\frac{-2\lambda_t^2}{(\hat\beta^2 - x_E^2)x_E}\,
\bigg\{
 \hat\beta x_E\,\big[-96 x_t^2 + 24 x_t x_H 
     - (-x_H + 1 + x_E)(x_E^2 - \hat\beta^2)\big] 
\nonumber\\[2mm] & & 
 {} +
   2(\hat\beta^2 - x_E^2)\,\big[ 24 x_t^2 + 2(x_H^2 - x_H x_E) 
   + x_t(-14 x_H + 12 x_E + x_E^2)\big]\,
   \ln\bigg(\frac{x_E + \hat\beta}{x_E - \hat\beta}\bigg) \bigg\}
.
\end{eqnarray}
These first two coefficients describe the s-channel exchange of the
photon and the Z boson where the Higgs boson is radiated 
off one of the top quarks~\cite{Dawson2}. A missing factor $s$ is
introduced in the first line of the formula for $G_2$. 

The coefficient functions $G_3$ to $G_6$ describe the emission of the Higgs 
boson from the Z-boson,
\begin{eqnarray}
G_3 & = &
\frac{ -2\hat\beta g_Z^2 x_t}{x_Z(x_H - x_Z + 1 - x_E)^2}
\,\bigg\{
    4x_H^2 + 12 x_Z^2 + 2x_Z x_E^2
\nonumber\\[2mm] & & 
 {} + (-1 + x_E)x_E^2 
    - x_H\big[ 8x_Z + (-4 + 4x_E + x_E^2)\big] \,\bigg\}
\,, \\[4mm]
G_4 & = & 
\frac{ \hat\beta g_Z^2 x_Z}{6(x_H - x_Z + 1 - x_E)^2}
\,\bigg\{
    48 x_t + 12 x_H - (-24 + \hat\beta^2 + 24 x_E - 3x_E^2) \,\bigg\}
\,, \\[4mm]
G_5 & = & 
\frac{4 \lambda_t g_Z x_t^{1/2}}{x_Z^{1/2}(-x_H + x_Z -1 + x_E)}
\,\bigg\{
   \hat\beta \big[ 6x_Z + x_E(-x_H - 1 + x_E)\big]
\nonumber\\[2mm] & & 
 {} +
    2\big[ x_H(x_H - 3x_Z + 1 - x_E) + 
         x_t(-4x_H + 12 x_Z + x_E^2)\big]\,
   \ln\bigg(\frac{x_E + \hat\beta}{x_E - \hat\beta}\bigg) \bigg\}
\,, \\[4mm]
G_6 & = & 
\frac{-8 \lambda_t g_Z (x_t x_Z)^{1/2}}{-x_H + x_Z - 1 + x_E}
\,\bigg\{
   \hat\beta + (4x_t - x_H + 2 - x_E) \,
   \ln\bigg(\frac{x_E + \hat\beta}{x_E - \hat\beta}\bigg) \bigg\}
\,.
\end{eqnarray}
These terms give contributions to the Higgs energy spectrum of less
than a few percent in the energy range between 500 GeV and 1 TeV.  
The overall signs of $G_5$ and $G_6$ are changed relative to
\cite{Dawson2}. The couplings and constants are defined in
Eqs.~(\ref{const1},\ref{const2}) and the term $\hat{\beta}$ is given by
\begin{equation}
\hat{\beta} \, = \, \left(\,
\frac{ 4\,(E_H^2 - m_H^2)\,(E_H^0 - E_H)}{\sqrt{s}\, 
(\,(E_H^0 - E_H)\,\sqrt{s} + 2m_t^2\,)} \,\right)^{1/2}
\,,
\label{betahut}
\end{equation}
with the large Higgs energy endpoint being defined as
\begin{equation}
E_H^0 \, \equiv \, \frac{ s+m_H^2-4 m_t^2}{2\sqrt{s}}
\,.
\end{equation}

\end{appendix}



\begin{thebibliography}{}


\bibitem{LEPlimits}
R.~Barate {\it et al.}  [ALEPH Collaboration],
Phys.\ Lett.\ B {\bf 565}, 61 (2003)
[arXiv:hep-ex/0306033].

\bibitem{MHupperlimit}
  M.~W.~Grunewald,
  arXiv:hep-ex/0511018.


\bibitem{TESLATDR}
J.~A.~Aguilar-Saavedra {\it et al.}  [ECFA/DESY LC Physics Working Group
                  Collaboration],
arXiv:hep-ph/0106315.

\bibitem{ALCphysics}
T.~Abe {\it et al.}  [American Linear Collider Working Group Collaboration],
in {\it Proc. of the APS/DPF/DPB Summer Study on the Future of Particle Physics (Snowmass 2001) } ed. N.~Graf,
arXiv:hep-ex/0106057;
%
T.~Abe {\it et al.}  [American Linear Collider Working Group Collaboration],
in {\it Proc. of the APS/DPF/DPB Summer Study on the Future of Particle Physics (Snowmass 2001) } ed. N.~Graf,
arXiv:hep-ex/0106056.

\bibitem{ACFALCphysics}
K.~Abe {\it et al.}  [ACFA Linear Collider Working Group Collaboration],
arXiv:hep-ph/0109166.


\bibitem{topmass}
T.~E.~W.~Group,
arXiv:hep-ex/0603039.
  

\bibitem{JustetopYukawa}
A.~Juste and G.~Merino,
arXiv:hep-ph/9910301.

\bibitem{GaytopYukawa}  
A.~Gay,
arXiv:hep-ph/0604034.

\bibitem{Hdecay}
A.~Djouadi, J.~Kalinowski and M.~Spira,
Comput.\ Phys.\ Commun.\  {\bf 108}, 56 (1998)
[arXiv:hep-ph/9704448].

\bibitem{Borneetth}
K.~J.~F.~Gaemers and G.~J.~Gounaris,
Phys.\ Lett.\ B {\bf 77}, 379 (1978);
%
A.~Djouadi, J.~Kalinowski and P.~M.~Zerwas,
Mod.\ Phys.\ Lett.\ A {\bf 7}, 1765 (1992);
%
A.~Djouadi, J.~Kalinowski and P.~M.~Zerwas,
Z.\ Phys.\ C {\bf 54}, 255 (1992).


\bibitem{Dittmaier1}
S.~Dittmaier, M.~Kramer, Y.~Liao, M.~Spira and P.~M.~Zerwas,
Phys.\ Lett.\ B {\bf 441}, 383 (1998)
[arXiv:hep-ph/9808433].


\bibitem{Dawson1}
S.~Dawson and L.~Reina,
Phys.\ Rev.\ D {\bf 57}, 5851 (1998)
[arXiv:hep-ph/9712400].


\bibitem{Dawson2}
S.~Dawson and L.~Reina,
Phys.\ Rev.\ D {\bf 59}, 054012 (1999)
[arXiv:hep-ph/9808443].


\bibitem{Belanger1}
G.~Belanger {\it et al.},
Phys.\ Lett.\ B {\bf 571}, 163 (2003)
[arXiv:hep-ph/0307029].


\bibitem{Denner1}
A.~Denner, S.~Dittmaier, M.~Roth and M.~M.~Weber,
Nucl.\ Phys.\ B {\bf 680}, 85 (2004)
[arXiv:hep-ph/0309274].


\bibitem{You1}
Y.~You, W.~G.~Ma, H.~Chen, R.~Y.~Zhang, S.~Yan-Bin and H.~S.~Hou,
Phys.\ Lett.\ B {\bf 571}, 85 (2003)
[arXiv:hep-ph/0306036].


\bibitem{FarrellHoang1}
  C.~Farrell and A.~H.~Hoang,
  Phys.\ Rev.\ D {\bf 72}, 014007 (2005)
  [arXiv:hep-ph/0504220].

\bibitem{HoangEpi}
  A.~H.~Hoang,
  Acta Phys.\ Polon.\ B {\bf 34}, 4491 (2003)
  [arXiv:hep-ph/0310301].

\bibitem{LMR} 
M.~Luke, A.~Manohar and I.~Rothstein,
Phys.\ Rev.\  {\bf D61}, 074025 (2000)
[arXiv:hep-ph/9910209].


\bibitem{HoangStewartultra}
A.~H.~Hoang and I.~W.~Stewart,
Phys.\ Rev.\ D {\bf 67}, 114020 (2003)
[arXiv:hep-ph/0209340].


\bibitem{hmst}
A.~H.~Hoang, A.~V.~Manohar, I.~W.~Stewart and T.~Teubner,
Phys.\ Rev.\ Lett.\  {\bf 86}, 1951 (2001)
[arXiv:hep-ph/0011254];
%
and Phys.\ Rev.\ D {\bf 65}, 014014 (2002)
[arXiv:hep-ph/0107144].


\bibitem{Justetalk}
A.~Juste talk presented at the Chicago Linear Collider Workshop, Chicago, USA,
January 2002, \url{http://www.pas.rochester.edu/~orr/justelc.pdf}. 


\bibitem{topQCDSnowmass}
  A.~Juste {\it et al.},
  arXiv:hep-ph/0601112;
%
  A.~Juste,
  arXiv:hep-ph/0512246.



\bibitem{Jezabek1}
M.~Je\.zabek, J.~H.~K\"uhn and T.~Teubner,
Z.\ Phys.\  {\bf C56}, 653 (1992).
%


\bibitem{Strassler1}
M.~J.~Strassler and M.~E.~Peskin,
Phys.\ Rev.\  {\bf D43}, 1500 (1991).



\bibitem{HoangReisser1}
A.~H.~Hoang and C.~J.~Reisser,
arXiv:hep-ph/0412258.


\bibitem{Hoangupsilon}
A.~H.~Hoang, Z.~Ligeti and A.~V.~Manohar,
Phys.\ Rev.\ Lett.\  {\bf 82}, 277 (1999)
[arXiv:hep-ph/9809423];
%
A.~H.~Hoang, Z.~Ligeti and A.~V.~Manohar,
Phys.\ Rev.\ D {\bf 59}, 074017 (1999)
[arXiv:hep-ph/9811239].


\bibitem{HoangTeubner}
  A.~H.~Hoang and T.~Teubner,
  Phys.\ Rev.\ D {\bf 60}, 114027 (1999)
  [arXiv:hep-ph/9904468].

\bibitem{Vcrenormalon}
A.~H.~Hoang, M.~C.~Smith, T.~Stelzer and S.~Willenbrock,
Phys.\ Rev.\ D {\bf 59}, 114014 (1999)
[arXiv:hep-ph/9804227];
%
M.~Beneke,
Phys.\ Lett.\ B {\bf 434}, 115 (1998)
[arXiv:hep-ph/9804241];
%
A.~Pineda, PhD Thesis, Univ. Barcelona (1998).


\bibitem{synopsis}
  A.~H.~Hoang {\it et al.},
  Eur.\ Phys.\ J.\ directC {\bf 2}, 1 (2000)
  [arXiv:hep-ph/0001286].



\bibitem{Parke1}
  S.~J.~Parke and Y.~Shadmi,
  Phys.\ Lett.\ B {\bf 387}, 199 (1996)
  [arXiv:hep-ph/9606419].


\bibitem{Pineda1}
A.~Pineda,
Phys.\ Rev.\ D {\bf 66}, 054022 (2002)
[arXiv:hep-ph/0110216].






\end{thebibliography}
\end{document}